\documentclass[12pt,a4paper,final]{iopart}

\usepackage{iopams}  
\usepackage{graphicx}
\usepackage{epstopdf}
\usepackage[breaklinks=true,colorlinks=true,linkcolor=blue,urlcolor=blue,citecolor=blue]{hyperref}
\begin{document}

\topical{Atomically crafted spin lattices as model systems for quantum magnetism}
\author{A Spinelli, M P Rebergen and A F Otte}
\address{Department of Quantum Nanoscience, Kavli Institute of Nanoscience\\ Delft University of Technology, Lorentzweg 1, 2628~CJ Delft, The Netherlands}
\ead{a.f.otte@tudelft.nl}

\begin{abstract}
Low-dimensional quantum magnetism presents a seemingly unlimited source of rich, intriguing physics. Yet, as realistic experimental representations are hard to come by, the field remains predominantly theoretical. In recent years, artificial spin structures built through manipulation of magnetic atoms in a scanning tunnelling microscope have developed into a promising testing ground for experimental verification of theoretical models. Here we present an overview of available tools and discuss recent achievements as well as future avenues. Moreover, we show new observations on magnetic switching in a bistable bit that can be used to extrapolate information on the magnetisation of the microscope tip.
\end{abstract}

%\submitto{\JPCM}
%\maketitle
\vspace{-0.5cm}
\tableofcontents

\newpage

\section{Introduction}
Magnetism is a collective phenomenon. Whereas a single spin is known to behave perfectly quantum mechanically, classical magnetic stability sets in only when many spins are coupled together. In between these two regimes there is a size domain -- anywhere between two and, say, hundreds of spins -- in which the system is large enough to speak of collective spin excitations, but still small enough in order for these excitations to be coherent throughout the system. It is in this region that we can learn how the collective quantum mechanical behaviour of interacting spins eventually gives rise to the emergence of the magnetic properties found in materials on the macroscopic scale.

Atomic assembly of magnetic structures through low temperature scanning tunnelling microscopy (STM) poses an intriguing method for entering this size domain. Less than 25~years after the first demonstration of atom manipulation \cite{Eigler1990}, we have now come to a time where it is possible not only to detect a single atom's magnetization \cite{Heinrich2004,Meier2008}, but also to analyse and control its magnetocrystalline anisotropy \cite{Hirjibehedin2007,Bryant2013,KhajetooriansPRL2013,Oberg2014}, measure its Kondo coupling to the substrate electrons \cite{Madhavan1998,Wahl2004,Neel2007,Otte2008}, tune and read-out interatomic exchange coupling between two neighbouring atoms \cite{Wahl2007,Otte2009,Zhou2010} and build extended atomic spin chains and structures \cite{Hirjibehedin2006,Neel2011,Khajetoorians2012}. Even the first steps towards technological usage of these abilities are taken in the form of an atomic scale logic gate \cite{KhajetooriansScience2011} and of atom-based bits showing extended magnetic lifetimes \cite{Loth2012,KhajetooriansScience2013,Miyamachi2013,Spinelli2014}. We believe that these techniques can be equally valuable for the more fundamental challenge of exploring quantum magnetism.

In this review we will focus on experiments performed on magnetic atoms deposited on thin insulating layers, specifically Cu$_2$N on Cu(100). This surface provides significant decoupling between the atoms and the bulk electrons in the metal below, resulting in relatively long lived spin states. The magnetic atoms we will focus on are $3d$-metals, in particular Fe and Co. Fe on Cu$_2$N has a spin $S=2$ and its ground state lies in the subspace with highest spin projection ($m=\pm2$). On the other hand, Co has spin $S=3/2$ and the ground state is in the multiplet with smallest spin projection ($m=\pm1/2$). As we will see in the next sections, those two magnetic atoms have very different properties, not only due to the difference between integer and half-integer spins. On this surface, they develop a different magnetocrystalline anisotropy, that gives rise to dissimilar behaviours both for a single atom and for structures composed of few atoms. Moreover, Co turns out to be Kondo-screened while Fe is not. In addition to reviewing existing work we will present novel results on the back-action of the atomic spin structures onto the magnetic state of the functionalized spin-polarized tip that is used to probe them. 

\section{Motivation: low dimensional quantum magnetism}
Inside a magnetically ordered material, all the atomic moments point in a specific direction: once the magnetization direction of one atom is fixed, the $n$th atom's magnetization will have a specific orientation, determined by the exchange interaction between neighbouring atoms. However, since the macroscopic magnetic state results from the interactions between all the constituent atomic spins, the emergence of magnetic order is very difficult to predict. In some materials magnetic order does not set in at all, regardless of system size. An interesting approach to gain insight into these processes is to study interactions on the scale of individual spins in a finite system, while slowly increasing the number of participating spins.

The two best known ordered materials are ferromagnets (FM) and antiferromagnets (AFM). In a ferromagnet, the lowest energy state is one where all spins are parallel to each other, giving rise to a net magnetization that is preserved even when the external magnetic field $\mathbf{B}=0$. In an antiferromagnetic material, in the ground state the spins exhibit antiparallel alignment, which gives rise to a zero net magnetization even though the magnetic order is present.

\begin{figure}[htb!]
 \centering
 \includegraphics{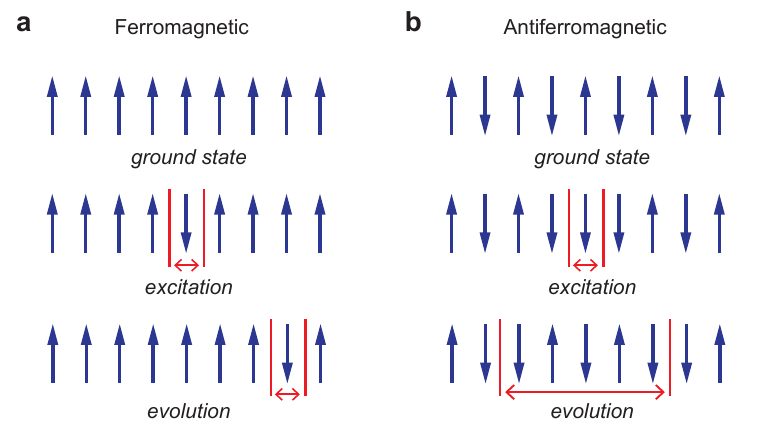}
 \caption{\label{fig:FMvsAFM}(a) Magnon excitation and its evolution inside a ferromagnetic chain. In the ground state all the spins are in parallel alignment within each other. A magnon can be represented as a single reversed spin surrounded by two domain walls. It can propagate along the chain, but the domain walls will stay bounded to each other. (b) Same as (a) but for an antiferromagnetic chain, in the Ising limit. In the ground state, the spins are in antiparallel alignment. An excitation is achieved by creating two domain walls that separate two different AFM phases. Those two fractional excitations can independently propagate along the chain.}
\end{figure}

In one dimension, the collective excitations from the ground state have a different nature for FM and AFM spin structures, as depicted in \fref{fig:FMvsAFM}. In a ferromagnetic phase, the excited states have the form of \textit{magnons}: single spin flipped by $\Delta m=1$ surrounded by two bound domain walls (\fref{fig:FMvsAFM}a). These domain walls are delocalised inside the structure and always propagate in pairs. The classical analogous of a magnon can be visualized as a spin wave, a coherent precession of the local spin expectation value around the common magnetization direction \cite{Mourigal2013}. In an antiferromagnet, the elementary excitations are \textit{spinons}. In this case the single flipped spin is surrounded by two domain walls that are not bound and can propagate independently. Each spinon carries a net spin value of 1/2 (\fref{fig:FMvsAFM}b). These fractional quasiparticles are easier to visualize in the Ising limit \cite{Ising,Brush1967}, in which all the spins exist only in two possible orientations. In this case, a spinon is simply a domain wall between two different antiferromagnetic phases. In the more complex Heisenberg case (which will be discussed in \sref{section:Heisenberg}) a single spin flip needs to be visualized with more pairs of spinons \cite{Mourigal2013}.     

When considering quantum spins, we may enter some regimes in which things can get more complex and the system can host exotic excitations and lose long range order. In particular, the system can become a quantum spin liquid (QSL), a spin system that is supposed to hold many exotic properties and has attracted massive theoretical interest after its first proposal, made by Anderson in 1973 \cite{Anderson1973}. A QSL is a collective spin state whose magnetic order is suppressed due to quantum fluctuations, leading to liquid-like properties among the spins, and it can emerge in presence of frustration or in low dimensions \cite{Ramirez2008,Balents2010}. 

The type of ground state and the nature of the excitations of a coupled spin system strongly depend on several parameters \cite{mila2000}: the magnitude of the spin, the interaction strength (we already mentioned the profound differences between FM and AFM coupling and the importance of long range order for the emergence of classical stability) and the geometry of the system. The magnitude of the spin is what makes it quantum: the case $S\rightarrow\infty$ corresponds to the classical limit, while $S=1/2$ is the extreme quantum limit. Also integer and half-integer spins are predicted to show quite different behaviours. For instance, while an antiferromagnetic chain composed of half integer spins has a relatively long range ordered ground state, the same chain composed of integer spins is predicted to lose this order even at zero temperature (the Haldane model \cite{Affleck1999}). The dimensionality of the system plays a critical role: in one- and two-dimensions there is no long range magnetic order, as stated in the Mermin--Wagner theorem\cite{MerminWagner}. Finally, quantum fluctuations in a spin system can be enhanced in presence of frustration, which implies a conflict in minimizing the coupling energies associated with different spin pairs; these conflicts arise mainly because of the lattice topology or due to the presence of competing further-neighbour interactions \cite{Bose2005}. 

The literature on quantum magnetism in general and on quantum spin liquids in particular is quite extensive; for a review see for instance Refs.~\cite{Lecheminant2003} and \cite{Intro_frustrated_magnetism}. Yet, despite the large interest in this topic from a theoretical viewpoint, the possibilities for experimental verification remain limited. On the one hand, magnetic materials are studied from the top down, using spatially averaging techniques such as neutron diffraction \cite{Mourigal2013}. While offering highly detailed information in reciprocal space, these techniques lack the possibility to probe materials locally or as function of system size. A different approach is provided by the community of ultracold atoms, where optical lattices containing spins are built from the bottom up \cite{Bloch2012}. The opportunity to create spin lattices with atomic precision in a real condensed matter environment, as described in this review, offers a large range of new possibilities that complement those offered by existing methods.

\section{Measurement techniques}
The scanning tunnelling microscope (STM) was invented in 1981 by Roher and Binnig \cite{BinnigAPL1982,BinnigPRL1982} as a tool that, by using vacuum tunnelling in a controlled way, was able to image conductive surfaces with unprecedented atomic resolution. Even though the STM is most famous for its imaging capabilities, this instrument can also be used for various techniques needed for surface characterization. In particular, single atoms and molecules can be manipulated and their local electronic and magnetic properties can be investigated via inelastic electron tunnelling spectroscopy (IETS). In this section we will describe how atom manipulation and tunnelling spectroscopy work, specifically in the case of magnetic atoms on Cu$_2$N, and give insight into the possible models to reproduce measured spectra. Also, we will discuss spin-polarized measurements and how the magnetic state of an atom can be probed with a magnetic STM tip. 

\subsection{Atom manipulation}\label{sec_atom_manipulation}
The most common form of atom manipulation is lateral manipulation, where atoms are dragged along a flat metal surface. This is possible due to the relatively weak van der Waals binding of the atoms to the surface: when the metal tip is lowered towards the atom, due to the additional van der Waals attraction of the tip the atom may fall into the combined potential well of both tip and surface, the minimum of which is located above the atom's original position. Combinations of materials that are often used for this technique are Fe atoms on Cu(111) (which is known for the quantum corral \cite{Crommie1993} and the atomic OR gate \cite{KhajetooriansScience2011}), CO molecules on Cu(111) (in the molecule cascades \cite{Heinrich2002} and `molecular graphene' \cite{Gomes2012}), and Fe atoms on Pt(111) \cite{KhajetooriansPRL2013}. The $\{111\}$ crystal orientation is particularly suitable for lateral atom manipulation due to its dense packing.

\begin{figure}[htb!]
 \centering
 \includegraphics{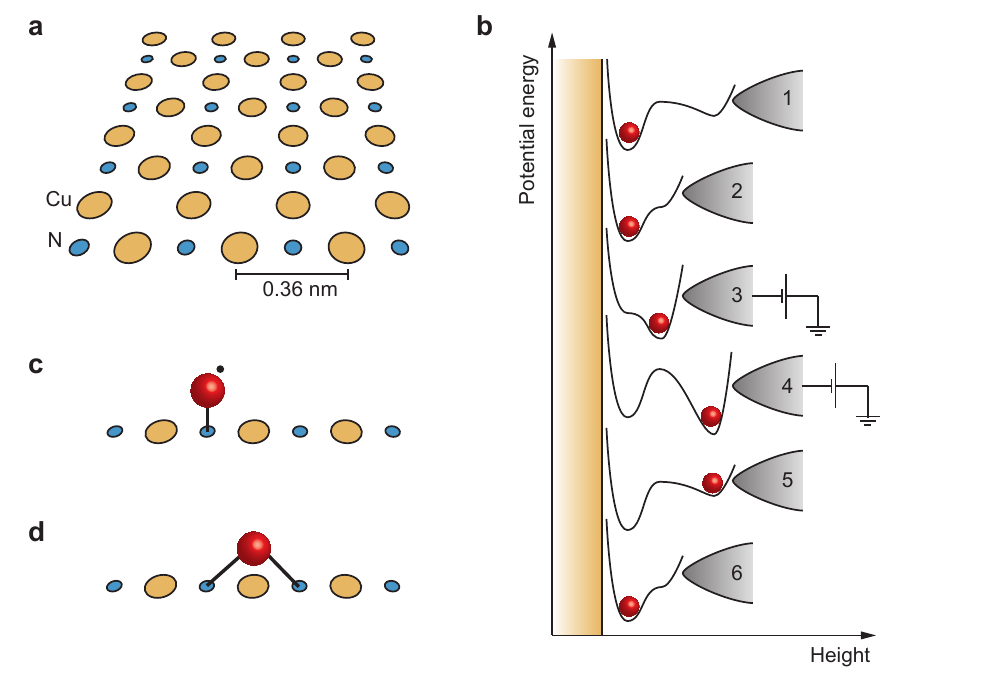}
\caption{\label{fig:manip}(a) Schematic showing the Cu$_2$N lattice: the orange circles represent Cu atoms while the blue circles are N atoms. The presence of strain on the underlying Cu(100) surface interferes in the growth process of the Cu$_2$N layer, that grows in islands, with typical sizes varying between 5~nm$^2$ and 20~nm$^2$\cite{Leibsle1993,Oberg2014}. (b) Visualization of pick-up and drop-off procedures for vertical atom manipulation. In all drawings, the left side of the double potential well represents the surface, the right side the tip. In (1) the tip is at imaging distance: the magnetic atom (red sphere) is stable on the surface. If the tip gets closer, the potential well is distorted (2). By applying a negative tip bias (the magnetic atom is positively charged), the right side of the potential well becomes more stable than the surface side and the atoms falls in (3). When the tip is then pulled away, the atom will move as well, as seen in (4). After restoring the imaging settings (5), the atom will stay on the tip side. When the tip is then moved closer to the surface, the atom will easily fall back to the surface (6). (c)-(d) Diagrams of the hopping process. In (c), the magnetic atom after being dropped off from the tip, is lying on top of a N atom with which it forms one valence bond (black line), leaving one electron unpaired. In (d), the Fe atom has hopped onto an adjacent Cu, and it is now forming two covalent bonds with the neighbouring N atoms. It is known from DFT calculations \cite{Hirjibehedin2007} that when an Fe atom is on top of a Cu, the latter is pushed down (lower than the neighbouring N atoms) and replaced in the lattice by the Fe.}
\end{figure}

Lateral manipulation does not work on the Cu(100)-$c(2\times2)$N surface \cite{Leibsle1993}, hence referred to as Cu$_2$N, the two-dimensional lattice which is shown in \fref{fig:manip}a. According to density functional theory (DFT) calculations, the covalent Cu--N bonds form a fishnet-shaped two-dimensional network on the surface \cite{Hirjibehedin2007}. As confirmed by STM topography, Fe atoms, upon being evaporated onto the cold surface, bind on the Cu sites of this lattice. Here they replace the position inside the covalent network previously held by the underlying Cu atom such that the resulting local structure can be seen as a N--Fe--N molecule. The same behaviour as for Fe was found for Mn and Co atoms \cite{Hirjibehedin2006,Otte2008}.

Since $3d$-metals tend to be strongly bound to the Cu$_2$N surface, a different technique is needed to move atoms, commonly called vertical manipulation \cite{Eigler1991,Hirjibehedin2006}. This technique is based on the actual transfer of the atom from the surface to the tip, and then again to the surface, allowing for an atomic precision drop-off. When the tip is few {\aa}ngstr{\"o}ms away from the surface, at a common imaging distance, the system can be seen as an asymmetric double potential well: the atom has two possible stable positions, one on the surface and one on the tip, that are separated by an energy barrier (\fref{fig:manip}b), with global minimum on the surface side. By letting the tip closer to the atom, and applying a negative voltage to the tip, the potential well gets distorted favouring the tip side of the junction, so that when the tip is lifted up the atom follows \cite{Hla2005, Ghosh2002}. This whole process is possible due to the charging effect induced by the lattice on the magnetic atom: in the N--Fe--N bond, negative charges move away from the Fe atom towards the neighbouring N atoms, rendering the Fe atom positively charged. On the other hand, when the tip with an Fe atom at its apex is brought close to the surface, the shape of the potential well will favour the atom to move on the surface, even if only a very small positive tip voltage (or none at all) is applied.

When the magnetic atom is dropped off from the tip, the minimum energy configuration is for it to land on a N site of the Cu$_2$N lattice, forming a single covalent bond. However, in this way, one electron remains unpaired (\fref{fig:manip}c). As soon as a voltage pulse of approximately $1.1$~V is applied to the Fe atom, it will move to an adjacent Cu site. In this new position, it will replace the Cu in the two-dimensional surface network by pushing it down, and will form two covalent bonds with the neighbouring N atoms (\fref{fig:manip}d). In this new position, it is so strongly bound to the surface that the only way to move it again is via a new pick up. If, while applying the voltage pulse the tip is slightly offset away from the center of the atom in the direction of one of the four neighbouring copper atoms, the magnetic atom can be forced to jump in that particular direction, with reasonable reliability.

\subsection{Inelastic electron tunnelling spectroscopy}\label{sec:IETS}
The primary tunnelling process occurring in an STM configuration is elastic tunnelling, where no energy is lost by the electrons. But it is also possible for the electrons to tunnel inelastically. In this process, electrons give away energy during the tunnelling process leading to an additional transport path in addition to the regular elastic tunnelling path. This energy is absorbed by the environment in the form of e.g. a vibrational \cite{Jaklevic1966,Stipe1998} or magnetic \cite{Heinrich2004} excitation. Since this excitation can occur only if the applied voltage exceeds the excitation energy, $eV>\Delta E$, the inelastic tunnelling path contributes to the conductance only beyond the excitation threshold. As a result, an inelastic excitation is seen in the differential conductance d$I$/d$V$ as a sudden upward step at the onset voltage.

\begin{figure}[htb!]
 \centering
 \includegraphics{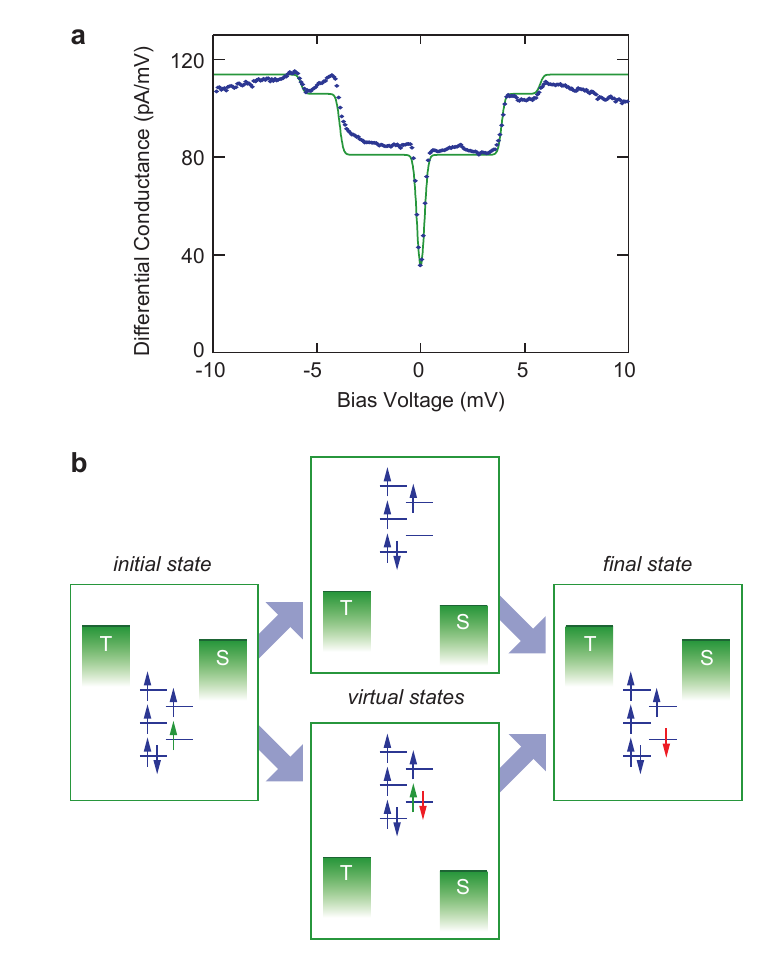}
 \caption{\label{fig:IETS}(a) d$I$/d$V$ spectrum (dark blue), and correspondent simulated curve (green) of a single Fe atom on Cu$_2$N/Cu(100), in absence of external magnetic field and at 330 mK. The steps are the inelastic spin excitations from the ground state. (b) Schematic of an inelastic cotunnelling process behind an STM measurement. In the initial configuration, the magnetic atom is in the ground state; in the final configuration, one spin has been flipped. This can happen in two possible virtual ways: a hole (a) or an electron (b) can tunnel to the magnetic atom without conserving the energy (the magnetic state is now outside the energy range of the two sides of the tunnel junction), and at the same time another hole or electron can tunnel from the magnetic atom to the other side of the barrier. In this way the system has changed its magnetization. The reason why it is possible to have this process, is that the time spent from the system in the virtual state (without energy conservation) is the Heisenberg uncertainty time $t_{\rm H}\simeq \hbar/E_{\rm C}$, with $E_{\rm C}$ the energy that the system should have paid to have a real excitation to the intermediate state \cite{Nazarov2009,Delgado2011}.}
\end{figure}

Differential conductance curves are commonly recorded at low temperature using lock-in detection at a frequency near bandwidth of the current amplifier ($\sim1$~kHz). The amplitude of the modulation signal gives an upper bound on the achievable energy resolution. \Fref{fig:IETS}a shows a typical IETS spectrum measured on an Fe atom on Cu$_2$N. Apart from the onset energies for each of the excitations, in this case $<0.5$~meV, $\sim 4.0$~meV and $\sim 5.5$~meV, which can be read off directly from the step positions, much more information can be extracted from this spectrum.

The \emph{height} of each step gives the transition intensity for the corresponding excitation: it gives a measure for the quantum mechanical selection rules governing the spin excitations \cite{Hirjibehedin2007}. The intensity $I_{0\rightarrow n}$, of the transition from the ground state $|\psi_0,\sigma_0\rangle$ to the $n$th eigenstate $|\psi_n,\sigma_n\rangle$, can be expressed as \cite{Fernandez-Rossier2009,Persson2009,Fransson2009,Lorente2009,LothNature2010}:

\begin{equation}\label{eq:intensity}
I_{0\rightarrow n}=\left|\langle\psi_n,\sigma_{\rm f}|\left({\bf S}\cdot{\bf s}+u\right)|\psi_0,\sigma_{\rm i}\rangle\right|^2,
\end{equation}
   
where the inner product of the spins ${\bf S}$ of the atom and ${\bf s}$ of the tunnelling electron, acting on their combined state, ensures that angular momentum is conserved: since the difference between the projections of $\sigma_{\rm i}$ and $\sigma_{\rm f}$, the initial and final states of the tunnelling electron spin, can be no more than 1, neither can the difference between the ground state of the atomic spin $\psi_0$ and the targeted excited state $\psi_n$. This becomes immediately evident when we rewrite \eref{eq:intensity} as:

\begin{equation}
I_{0\rightarrow n}=\left|\langle\psi_n,\sigma_{\rm f}|\left(\frac{S_{+}s_{-}+S_{-}s_{+}}{2}+S_{z}s_{z}+u\right)|\psi_0,\sigma_{\rm i}\rangle\right|^2.
\end{equation}

Note that the $S_{z}s_{z}$ term also allows elastic contributions (i.e. where $\psi_{n}=\psi_{0}$). In addition, the constant $u$ represents an alternative elastic tunnelling path, possibly through entirely different orbitals. The two elastic terms are referred to as `spin-dependent' and `spin-independent' elastic tunnelling respectively.

When we want to compare two d$I$/d$V$ spectra taken on different atoms we need to normalize them; this is not a trivial issue. One could think of normalising them with current: every spectrum is recorded with the same initial current value at a specific bias voltage. However, the current is the integral of the d$I$/d$V$ spectrum that we want to measure, and therefore the normalization of the measurement would always be influenced by the measurement itself. Also normalizing with respect to the tip height over an atom is not appropriate for the same reason: the apparent height is still related to the current and therefore the integral of the d$I$/d$V$. To avoid these problems, we choose to normalize the measurement on the bare Cu$_2$N: we set current and voltage on the substrate and then, without changing the tip height that stays fixed from the off-atom value, we move the tip on the atom that we want to measure and we can record the d$I$/d$V$ spectra.

The \emph{shape} of a step beyond the onset voltage, as can be seen on the negative voltage side in \fref{fig:IETS}a, is not necessarily flat. Instead, in many cases the signal reaches a maximum after which it decays somewhat. Several causes can result in such nonlinear behaviour. First, saturation of the excited state may cause the inelastic signal to decline: beyond a certain voltage, above the excitation threshold, the population of the excited state becomes so high, and the population of the ground state so low, that gradually less is left to excite \cite{LothNature2010,Delgado2010}.

The simulated IETS spectrum of a single Fe atom on Cu$_2$N is shown in \fref{fig:IETS}a. The energies of the excitations follow directly from diagonalization of the total Hamiltonian of the system \eref{total_H} that will be derived in \sref{section:Heisenberg}. The full d$I$/d$V$ spectrum is modelled by using \cite{Lambe1968,Kogan2004,Hirjibehedin2007,OttePhD}

\begin{equation}\label{eq:sim_dI/dV}
\frac{{\rm d}I}{{\rm d}V}\propto\sum_n I_{0\rightarrow n} \times \left[F\left(\frac{eV+\Delta E_n}{k_{\rm B}T}\right)+F\left(-\frac{eV-\Delta E_n}{k_{\rm B}T}\right)\right],
\end{equation}

where the function $F$, defined as $F(x)=[1+(x-1)\exp{x}]/{[\exp{x}-1]}^2$, gives the correct shape to the IETS steps (taking into account thermal broadening), while $\Delta E_n$ is the energy difference between the $n$th and the ground state and $eV$ is the applied voltage. 

As can be seen in \fref{fig:IETS}a, the d$I$/d$V$ curve simulated with \eref{eq:sim_dI/dV} cannot reproduce the aforementioned nonlinear behaviour previously identified in the negative voltage side of the Fe spectrum. Whether saturation is the dominant reason for the nonlinear lineshape can be checked by reducing the total current (i.e. by increasing the tip height): when the (inelastic) current is smaller, saturation should be reached at a larger voltage and the decay becomes less steep. In this case, those lineshapes can be reproduced by taking into account also secondary transitions that start from an excited state. This can be done with the use of Pauli Master equations \cite{LothNature2010,Delgado2011}. In this method, the time evolution of every state is calculated taking into account the total transition rate into and out of that state, considering all possible spin contributions to the the total rate. Moreover, this saturation effect can be used to measure the lifetime of the excitation: by analysing the lineshape we can determine what magnitude of inelastic current is needed to compete with the relaxation process. With this method, the lifetimes of the excited states of individual Mn atoms on Cu$_2$N were found to vary between 0.1~ns and 1.0~ns \cite{LothNature2010}.

In other situations, nonlinearities in the IETS spectra are not caused by saturation and persist even for small currents.
In these cases, higher order terms need to be taken into account in the description of the tunnel current. Very similar lineshapes can be caused for instance by higher order effects due to the Kondo interaction of the atom with the bulk conduction electrons \cite{Zhang2013,Co-dimers}, as will be explained in \sref{section_Kondo}.

Finally, the \emph{width} of the steps is a measure for thermal (or modulation) broadening of the IETS signal or due to lifetime broadening of the spin states, whichever of the two is largest. Since spin lifetimes on Cu$_2$N are in the order of 0.1~ns or longer, measurements performed at typical temperatures of $\sim1$~K or higher will be dominated by thermal broadening. Spin excitation IETS has been observed in various other systems, including molecular magnets \cite{Chen2008,Tsukahara2009}, magnetic dopants in semiconductors \cite{Khajetoorians2010} and magnetic atoms on metal substrates \cite{KhajetooriansPRL2013}.

As a transport phenomenon, inelastic tunnelling is well described in terms of cotunnelling in the Coulomb blockade regime \cite{Delgado2011,Averin1990}: a framework that is used very successfully in the field of quantum transport through quantum dots \cite{DeFranceschi2001,Zumbuhl2004,Schleser2005}. In cotunnelling, the energy levels between which excitations are made are assumed to be well outside the energy window defined by the applied bias voltage (\fref{fig:IETS}b). Spin-flip excitations involve temporary virtual double occupancy (or de-occupancy) of orbitals that are singly occupied when at rest.

\subsection{Functionalized spin-polarized tips}\label{section_SP-STM}
The invention of the Scanning Tunnelling Microscope in the early 1980s has revolutionized the study of conductive material surfaces. However, the first proposal for implementing the STM to observe spin-contrast, and therefore atomic-scale magnetism, was produced only at the end of the decade by Pierce \cite{Pierce1988}. The principle of a spin-polarized (s-p) STM is to read the polarization of the surface spins with the tunnelling electrons provided by the tip, that have to be spin-polarized as well. The resulting conductance will depend on the overlap of the polarized electron states in both tip and sample.

The simplest of the methods proposed by Pierce was to create a polarized tunnelling current using a magnetic tip. The first spin-polarized STM experiment \cite{WiesendangerPRL1990} was performed with a ferromagnetic CrO$_2$ tip, that was used to get a topographic contrast of the antiferromagnetically ordered Cr terraces of a (100) surface. The apparent height of the atomic steps was different if the tip and the surface polarizations were parallel or not, allowing to obtain nanoscale magnetic resolution. The development of new methods to prepare sharper tips, allowed soon to achieve atomic spin contrast: in Ref.~\cite{WiesendangerScience1992} it is shown for the first time magnetic imaging at the atomic level, using a very sharp Fe tip to image a magnetite sample. After this improvement, the field of s-p STM has produced many studies of atomic scale magnetism using magnetic or magnetically coated tips, like e.g. Refs.~\cite{HeinzeScience2000, Bode2007, Gao2008} in the first decade of 2000 (other examples can be found in the review papers \cite{Bode2003,WiesendangerRev2009}), or more recently \cite{Sonntag2014, KhajetooriansScience2013}. 

Another method to achieve a spin polarized STM tip is to start with a nonmagnetic tip, and prepare it \textit{in situ} by transferring material from the magnetic surface to the tip, creating a magnetic cluster on the tip apex to make it sensitive to the spin orientation. One way of doing this is by applying voltage pulses \cite{Yamada2003}: this was shown to cause mass transport from sample to tip and to vary the magnetization direction of the STM tip. Alternatively, the tip can be indented into the magnetic substrate to obtain the same effective coating \cite{Berbil-Bautista2007}. These methods are easier to achieve than coating a non magnetic tip with a thin film of magnetic material \cite{WiesendangerRev2009}.

After the invention of vertical atom manipulation, this last technique to polarize the STM tip could be performed in a controlled way. It was demonstrated that if a single magnetic atom is picked up from the surface with a non magnetic tip and replaces the tip apex, the tip becomes a paramagnet \cite{OttePhD,LothNature2010}. When an external magnetic field $\mathbf{B}\sim100$~mT is applied, due to the Zeeman effect the tip will acquire a spin-polarization parallel to $\mathbf{B}$. This method enables to switch very easily from a non-magnetic tip to a s-p one, allowing to measure the same atom with or without spin contrast \cite{Spinelli2014}.  

Tersoff and Hamann \cite{Tersoff1983,Tersoff1985} have theoretically related the experimentally measured tunnel current and the electronic states of both tip and sample, starting from the perturbative approach introduced by Bardeen \cite{Bardeen1961}. Without going into mathematical details, that can be found for instance in Ref.~\cite{WiesendangerRev2009}, they found an exponential dependence of the current on the distance between tip and sample $\mathbf{d}$:

\begin{equation}\label{eq:I}
I_0(\mathbf{d}) \propto \exp{(-2\kappa \mathbf{d})}.
\end{equation}

This well-known relation is valid only for a non polarized tunnel junctions, since it is based on the assumption that the tip wavefunction is of the $s$-type. As soon as a tip polarization is introduced, the current needs to be corrected to properly take it into account. It can be expressed as \cite{WiesendangerRev2009}:

\begin{equation}
I_{SP}(\mathbf{d}) \propto I_0(\mathbf{d})\left[1+P^{T}P^{S}cos\beta\right],
\label{eq:I_SP}
\end{equation}

where $I_0(\mathbf{d})$ is the non polarized tunnel current \eref{eq:I} that exponentially depends on the distance $\mathbf{d}$, $\beta$ is the angle between tip and sample polarization directions, and $P^{T,S}$ are the polarizations of respectively tip and sample, defined as ($\alpha$ can be either tip or sample):

\begin{equation}
P^{\alpha}=\frac{n_{\uparrow}^{\alpha}-n_{\downarrow}^{\alpha}}{n_{\uparrow}^{\alpha}+n_{\downarrow}^{\alpha}}\label{eq:Polarization},
\end{equation}

with $n_{\uparrow}^{\alpha}$ and $n_{\downarrow}^{\alpha}$ being the spin-resolved densities of states for the two magnetic electrodes. 

\begin{figure}[htb!]
 \centering
 \includegraphics{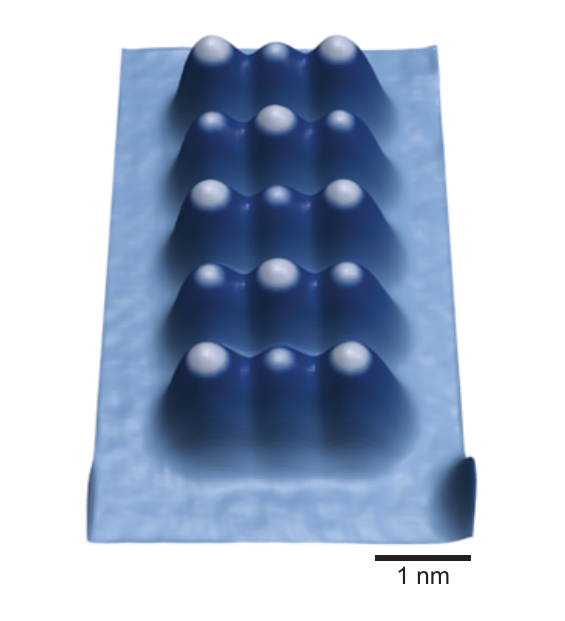}
 \caption{\label{fig:SP_structure} Antiferromagnetic structure composed of 15 Fe atoms on a Cu$_2$N/Cu(100) substrate, imaged with a spin polarized tip. A magnetic field $B=1$~T is applied in plane, parallel to the easy axis direction of the Fe atoms, to polarize the tip. The apparent height difference between the atoms is $\Delta y=15$~pm. This structure is similar to the antiferromagnetic arrays presented in Ref.~\cite{Loth2012}.}
\end{figure}

This difference in current between states with opposite magnetization directions, can be visualized as apparent tip height variation when imaged in constant-current mode. In \fref{fig:SP_structure} we show a topographic image of an antiferromagnetic structure composed of 15 Fe atoms on a Cu$_2$N lattice, with spacing similar to the structures presented in Ref.~\cite{Loth2012}. The apparent high-low pattern among the atoms is an indication of the antiferromagnetic coupling in the structure: since the tip is polarized in the same direction as the atoms on the surface, an atom that looks higher is magnetized in a parallel alignment to the tip apex, while an atom that looks lower is in an antiparallel configuration. We note that since the number of atoms in the structure is odd, the shown N{\'e}el state is stable. For an even numbered case, the stability depends on the number of coupled atoms \cite{Loth2012}, as will be explained in \sref{sec_bistability}. 

\section{Modelling the environment}
In the previous section we have seen how spin excitations appear as steps in an IETS measurement. Now we will focus on the physics of the spin states themselves. In our experiments on Cu$_2$N, the atomic spin's excitation energies are governed by three factors, all working on comparable energy scales: (i) magnetic anisotropy, (ii) the Zeeman effect and (iii) Heisenberg interaction with neighbouring atoms.

\subsection{Magnetic anisotropy}\label{section_anisotropy}
When a magnetic atom is embedded inside a crystal structure (or placed on top of a crystalline surface), broken symmetries between the different crystallographic axes may lead to some orientations of the magnetic moment being favoured over others: the primary ingredient for magnetic stability. This breaking of symmetry stems from the electrostatic potential of the crystal lattice having a different effect on each of the partially filled d-orbitals (for transition metals), in which the atom's net magnetic moment resides.

If this crystal field splitting is to such a degree that no degeneracy between any two d-orbitals remains, no electron can be in a state that has a considerable finite expectation value for the orbital angular momentum ${\bf L}$, for which always a linear combination of two orbitals is needed, and we say that the orbital momentum has \emph{quenched}. In such cases, the total magnetic moment of the atom is best described by ${\bf S}$, the combined spin momenta of the atom's uncompensated electrons. Whatever remains of ${\bf L}$ is felt by this spin ${\bf S}$ through the spin-orbit coupling $\lambda{\bf L}\cdot{\bf S}$ as a small perturbation ($\lambda$ is the spin-orbit constant). Depending on the orientation of ${\bf S}$ with respect to the orbitals this may lead to either an energy gain or loss, giving rise to magnetic anisotropy.

Traditionally, the magnetic anisotropy is quantified by the anisotropy Hamiltonian \cite{Gatteschi2006}

\begin{equation}\label{eq_anisotropy_DandE}
\mathcal{H}_{\rm aniso}=DS_z^2+E\left(S_x^2-S_y^2\right),
\end{equation}

where $D$ and $E$ are respectively the uniaxial and transverse anisotropy parameters. By definition, $D$ corresponds to the magnetic orientation whose anisotropy energy stands out most (either in a positive or a negative way), while $E$ gives the energy difference between the remaining two orientations. For example, if $D<0$ and $E>0$, it follows that $z$ is the easy axis, $y$ the intermediate axis and $x$ the hard axis.

In the ideal case where $E=0$ and $D<0$ (easy axis uniaxial anisotropy), the energy landscape of the spin takes the shape of a parabola with height $DS^2$ ($S$ being the magnitude of ${\bf S}$) separating magnetization eigenstates $|m_z\rangle=|{-}S\rangle$ and $|m_z\rangle=|S\rangle$. Since the $S_x$ and $S_y$ operators do not commute with $S_z$, the presence of a small transverse anisotropy $E$ will lead to mixing of the states. For an integer spin system this results in the two lowest states becoming symmetric and antisymmetric superpositions of $|{-}S\rangle$ and $|S\rangle$. In this case it is possible for the magnetization to tunnel from one orientation to the other and magnetic stability is lost. Only when a magnetic field $E/\mu_{B}\ll{\bf B}\sim DS^2/\mu_{B}$ ($\mu_B$ being the Bohr magneton) is applied along the $z$-direction do the states sufficiently separate again to form a bistable system; for higher fields, the state fully aligned to ${\bf B}$ becomes highly favourite.

For half-integer spins there is also mixing of states, but not within doublets due to Kramers' theorem \cite{Kramers}. This can be easily seen when we rewrite the anisotropy Hamiltonian in terms of the spin raising and lowering operators $S_{+}$ and $S_{-}$:

\begin{equation}
\mathcal{H}_{\rm aniso}=DS_z^2+\frac{E}{2}\left(S_{+}S_{+}+S_{-}S_{-}\right),
\end{equation}

which can only couple states whose $m_{z}$ values are a multiple of 2 apart. Instead, half-integer spins with transverse anisotropy can tunnel from e.g.\ $|{-}S\rangle$ to $|S-1\rangle$ or from $|{-}S+1\rangle$ to $|S\rangle$.

\begin{figure}[htb!]
 \centering
 \includegraphics{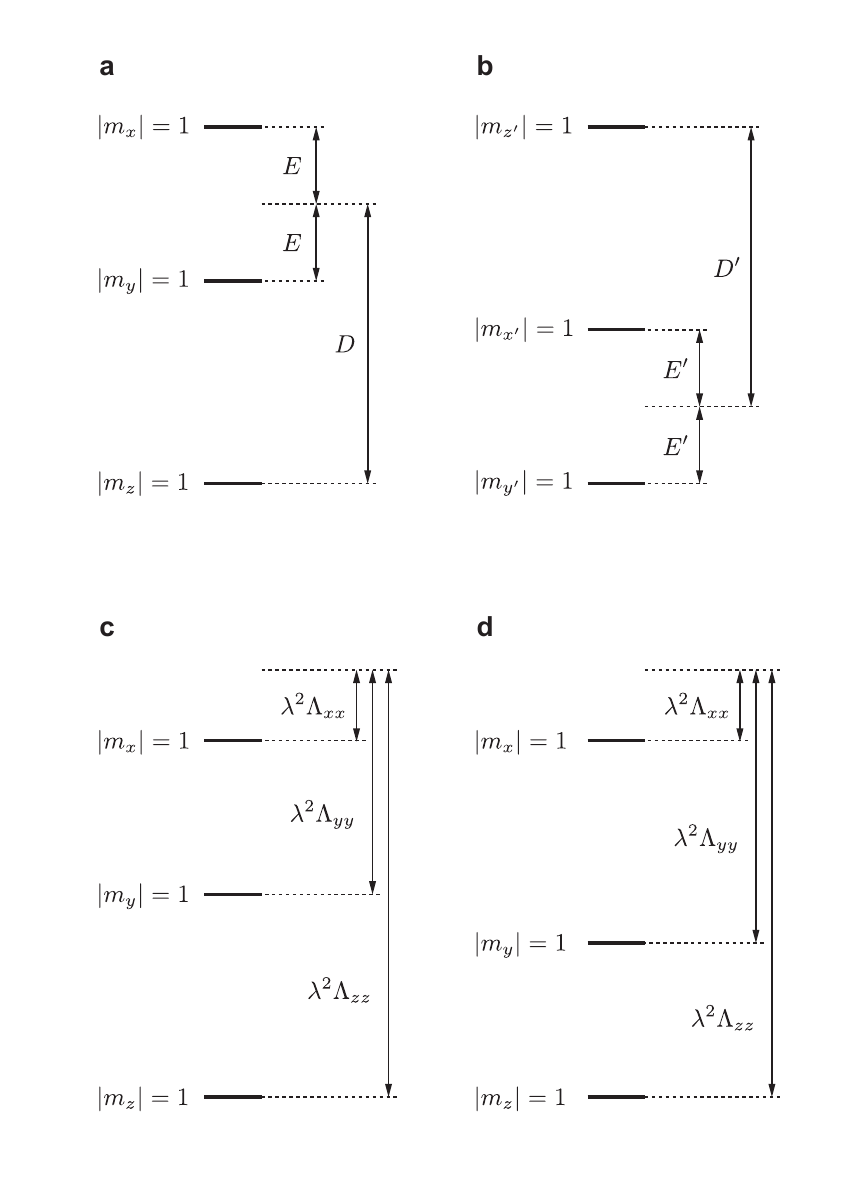}
 \caption{\label{fig:DEvsLambda}(a) Energy diagram for a $S=1$ system being fully aligned along the $x$, $y$ and $z$ axes, with easy axis anisotropy ($D<0$) and a transverse anisotropy $E$ which is just smaller than $|D|/3$. (b) Similar to (a), but with $E$ just larger than $|D|/3$, calling for new anisotropy parameters $D'$ and $E'$ as well as rearranged axes $x'$, $y'$ and $z'$. (c),(d) Same situations as in (a) and (b), but now represented in terms of $\Lambda$. The dashed line representing zero energy here is arbitrary: a random identical offset can be added to each of the $\Lambda_{\mu\mu}$ parameters as long as they all stay positive.}
\end{figure}

The representation in terms of $D$ and $E$ becomes inconvenient whenever $E$ approaches $|D|/3$. As shown in \fref{fig:DEvsLambda}a and \fref{fig:DEvsLambda}b, a small variation in the anisotropy energy of e.g. the $y$-orientation may lead to a situation where $D$ changes sign, switching from an easy axis to a hard axis configuration or vice versa, and all the axes need to be rearranged. One could decide not to invert $D$ and leave the axes unchanged (i.e. to abandon the requirement that $D$ always corresponds to the orientation that stands out most) but this would lead to the possibility of multiple combinations of $D$ and $E$ describing the same anisotropy configuration, which is even less desirable. A second disadvantage of this representation is that the $D$ and $E$ parameters are only phenomenological and do not provide much insight into the origin of anisotropy as a result of spin-orbit coupling.

For both these reasons, it is convenient to switch to a representation in terms of the anisotropy tensor $\Lambda$, whose elements are the matrix elements of the orbital momentum in second order perturbation theory \cite{Pryce1950}:

\begin{equation}
\Lambda_{ij}=\sum\limits_{n\neq 0}\frac{\langle\psi_0|L_i|\psi_n\rangle\langle \psi_n|L_j|\psi_0\rangle}{E_n-E_0}.
\end{equation}

Here the states $\psi_n$ refer to different orbital filling configurations with $\psi_0$ the ground state configuration. In a case of a high-symmetry crystal environment, all the off-diagonals of $\Lambda$ are typically zero, leaving only \cite{Dai2008}:

\begin{equation}\label{eq:Lambdas}
\Lambda_{\mu\mu}=\sum\limits_{n\neq 0}\frac{\left|\langle\psi_0|L_\mu|\psi_n\rangle\right|^2}{E_n-E_0},
\end{equation}

with $\mu=x,y,z$. In terms of these matrix elements the anisotropy Hamiltonian can be rewritten as:

\begin{equation}\label{eq_anisotropy}
\mathcal{H}_{\rm aniso}=-\lambda^2\left(\Lambda_{xx}S_x^2+\Lambda_{yy}S_y^2+\Lambda_{zz}S_z^2\right)\equiv -\lambda^2\sum_{\mu=x,y,z}\Lambda_{\mu\mu}S_{\mu}S_{\mu}.
\end{equation}

In this form the Hamiltonian is very easy to read: whichever $\Lambda_{\mu\mu}$ is largest simply corresponds to the easy axis while the smallest $\Lambda_{\mu\mu}$ corresponds to the hard axis. In addition, as shown in figures~\ref{fig:DEvsLambda}c--d, there is no longer the need to rearrange the axes when the system switches between easy axis and hard axis configurations.

One could say that $\Lambda_{\mu\mu}$ represents the extent to which $L_{\mu}$, the component of the orbital angular momentum in the $\mu$-direction, is \emph{un}quenched. Of course, this quantity can never exceed $L$, the magnitude of ${\bf L}$. When the energy difference between two states $\psi_n$ becomes too small, $\Lambda_{\mu\mu}$ will rapidly increase and the approach of treating the spin-orbit coupling as a perturbation breaks down. In this case the anisotropy energy can be read off directly from the $\lambda{\bf L}\cdot{\bf S}$ term in the Hamiltonian. From this it can be seen that the maximum height of the anisotropy barrier, i.e. the maximum difference in energy between to perpendicular orientations of ${\bf S}$, can never exceed $\lambda LS$.

\subsubsection{Strain induced variation of magnetic anisotropy}

The Hamiltonian \eref{eq_anisotropy} now expresses the magnetic anisotropy in terms of physical parameters. Using this notation makes it is easier to understand the significant variations in anisotropy measured in some cases. In Ref.~\cite{Bryant2013}, three different types of Fe dimers were built via vertical manipulation. These pairs of atoms were chosen such that they all have the same separation distance in terms of Fe--N and Cu--N bonds, but their precise geometries were different. Inelastic tunnelling spectra were recorded on every atom of the dimer: for pairs in which the second atom was along (or within one bond length of) the easy axis of the first atom, an enhancement of the magnetic anisotropy, with respect to the isolated Fe atom value, was measured. On the other hand, when the second atom was further away from the easy axis, a decrease in the magnetic anisotropy was found.

\begin{figure}[htb!]
 \centering
 \includegraphics{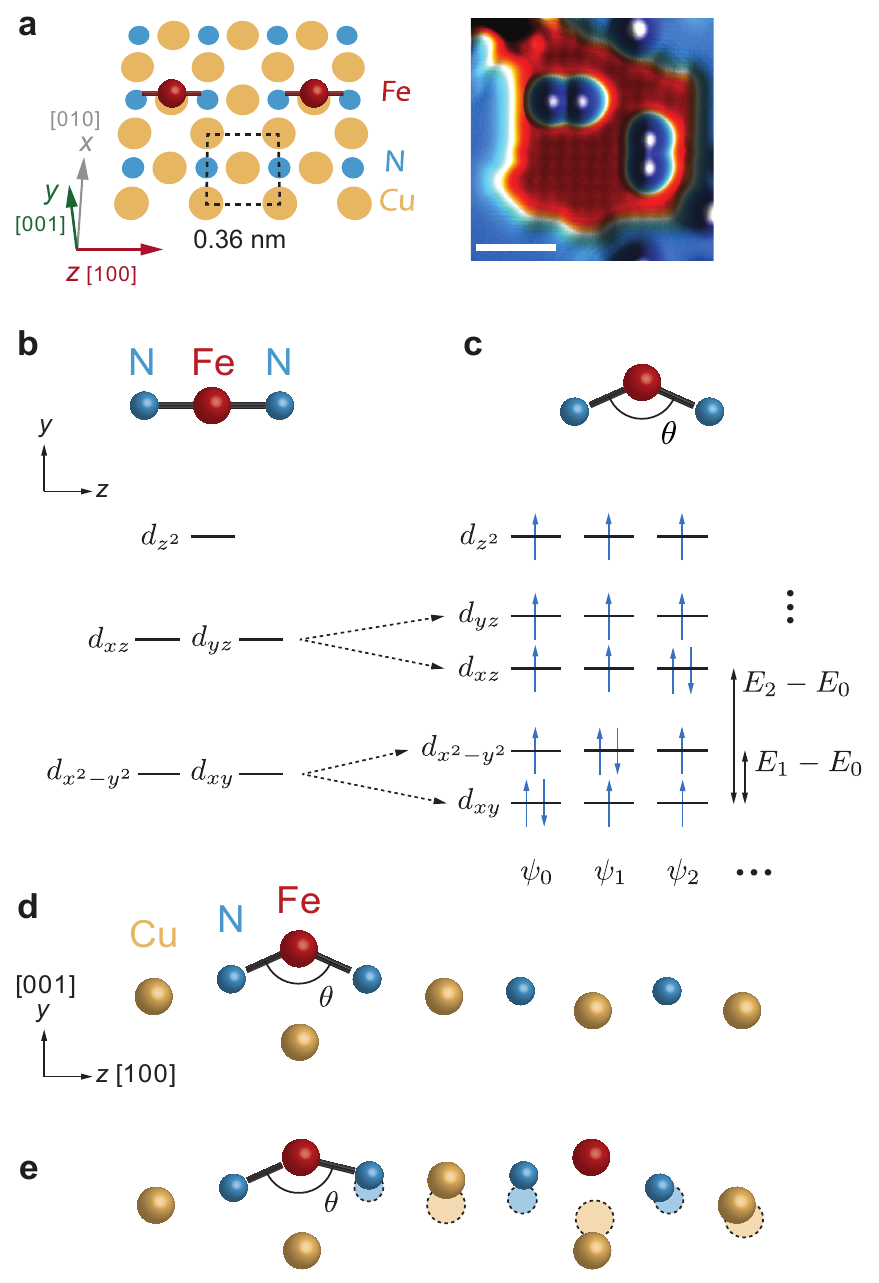}
\caption{\label{fig:Fe_dimer}(a) Diagram and STM topographic image of two instances of the linear dimer. Red bars indicate the magnetic easy axis along a nitrogen row. The unit cell of Cu$_2$N is also shown. Scale bar in topography is $2$~nm. (b) Proposed splitting of the $d$-orbitals resulting from a linear field. (c) Same as (b) but for a $C_{2v}$ crystal field ($\theta<180^{\circ}$). Now all the orbital degeneracies have been lifted. The orbital configurations for the three lowest energy orbital states $\psi_n$ are shown. (d) Schematic of a single Fe on Cu$_2$N. (e) Same as (d) but for the linear dimer: the presence of the second atom lifts the central Cu atom, in this way increasing $\theta$. The dashed outlines show the corresponding lattice displacement in the case of a single Fe atom. All panels edited from Ref.~\cite{Bryant2013}.}
\end{figure}

A model was developed to explain the observed results, based on the action of strain induced by the presence of the second atom, that changes the local crystal field. Of the three different instances of dimers described in Ref.~\cite{Bryant2013}, only the more instructive one, shown in \fref{fig:Fe_dimer}a, will be discussed here. This is the building block of the magnetic bistable structures presented in Ref.~\cite{Loth2012}; the two Fe atoms are positioned along a N-row of the Cu$_2$N network, spaced two unit cells ($0.72$~nm apart), and for both atoms a significant increase of the magnetic anisotropy is detected.

As seen in \sref{sec_atom_manipulation}, a single Fe atom on Cu$_2$N binds on a Cu site. According to DFT calculations the Fe atom pushes this Cu atom down into the bulk and replaces its position in the surface molecular network \cite{Hirjibehedin2007}. The Fe atom forms two covalent bonds with the neighbouring N atoms, that are slightly pushed downwards as well, thus constituting a N--Fe--N structure. If this structure is linear, the corresponding crystal field is linear as well, and the produced splitting of the Fe $d$-orbitals is shown in \fref{fig:Fe_dimer}b. However the crystal field created by this environment is not linear, but has a $C_{2v}$ symmetry, caused by the presence of the angle $\theta$ in the N--Fe--N bond. The result of this symmetry is that all the degeneracies between the orbital eigenstates are lifted (\fref{fig:Fe_dimer}c). 

The anisotropy parameters $\Lambda_{\mu\mu}$, as defined in \eref{eq:Lambdas}, are inversely proportional to the energy difference between those states. In particular, $\Lambda_{zz}$, the amount of unquenched orbital momentum along the nitrogen direction of the lattice ($z$-direction), is inversely proportional to $E_1-E_0$, the energy difference between the ground state and the first orbital excited state. This energy difference will vanish in the limit of $\theta=180^{\circ}$ in the N--Fe--N bond. The other two components of the anisotropy tensor will depend on the energy differences $E_2-E_0$ and $E_3-E_0$, that are less strongly influenced by the angle $\theta$.

In \fref{fig:Fe_dimer}d a schematic, based on DFT calculations, of a single Fe atom embedded inside a Cu$_2$N lattice is shown: the neighbouring N atoms and the next-nearest-neighbour Cu atoms are lifted and move toward the Fe atom. When a second Fe atom is placed along the same lattice row, it will lift the neighbouring N atoms, like the first Fe (\fref{fig:Fe_dimer}e), and the central Cu atom will be raised more, since it is now interacting with both Fe atoms. As a consequence, the angle $\theta$ will increase, resulting in an enhancement of $\Lambda_{zz}$. To summarize, the presence of the second atom has induced strain on the lattice that has modified the local crystal field felt by the single Fe atom, and therefore the splitting within its $d$-orbitals, leading to a significant variation of the magnetic anisotropy.  

Analogous reasoning can be made to explain all observed variations between different dimer geometries. The dimer discussed here, however, since it is built along a nitrogen row of the lattice, is the one most sensitive to variations in strain: it is in a critical regime in which even the smallest modification of the angle $\theta$ can induce a very large variation in the magnetocrystalline anisotropy. Comparable results were found for Co pairs on Cu$_2$N \cite{Co-dimers}. Also in this case, the dimer built with the same geometry was the most sensitive to strain and produced the largest variation in anisotropy compared to isolated atom value.

\subsection{Heisenberg interaction}\label{section:Heisenberg}
If two or more magnetic atoms are placed on a surface at a close distance, their electrons will interact. This interaction can be direct, like in the formation of a covalent bond \cite{Griffiths_QM}, but it can also be indirect: mediated by a non magnetic atom, commonly known as superexchange interaction \cite{Koch2012}, or mediated by the delocalised electrons of a conducting material (Ruderman-Kittel-Kasuya-Yosida or simply RKKY interaction) \cite{Ruderman1954,Kasuya1956,Yosida1957}. Both these interactions are collinear, favouring either parallel or antiparallel alignment of the spins. Until now, no clear picture as emerged as to which of the two interactions dominates for magnetic atoms on Cu$_2$N. In none of the experiments evidence of non-collinear interaction, such as Dzyaloshinskii-Moriya as found in magnetic surfaces \cite{Bode2007}, have been reported.

The strength of the interaction can be tuned by changing the relative positioning of the atoms on the lattice \cite{Bryant2013}. We can model their magnetic interaction using the Heisenberg exchange Hamiltonian \cite{Fernandez-Rossier2009}:

\begin{equation}\label{eq:Heisenberg_int}
\mathcal{H}_{\rm Heis}= \sum_{i,j} J_{ij} {\bf S}_i \cdot {\bf S}_{j},
\end{equation} 

with the sign of $J$ defining whether the coupling is ferromagnetic (FM, $J<0$) or antiferromagnetic (AFM, $J>0$). In our system we find that it is generally sufficient to consider only interactions between nearest neighbours, therefore limiting the sum to $j=i+1$.

If we consider all the different contributions, we can write the total Hamiltonian for a system composed of $N$ magnetic atoms on a surface, coupled within nearest neighbours with strength $J$ and in presence of an external magnetic field ${\bf B}$ ($\mu_{\rm B}$ being the Bohr magneton):

\begin{equation}\label{total_H}
\mathcal{H}=J{\bf S}^{i}\cdot{\bf S}^{(i+1)}-\sum_{i,\mu}\left[\lambda^2\Lambda_{\mu\mu}S_{\mu}^{i}S_{\mu}^{i}+2(1-\lambda\Lambda_{\mu\mu})\mu_{\rm B} B_{\mu}S_{\mu}^{i}\right],  
\end{equation} 

where the index $i$ refers to the $i$-th atom in the structure. We point out that when expressing the spin-orbit interaction in terms of the $\Lambda_{\mu\mu}$ parameters, the $g$-factor becomes a tensor \cite{Pryce1950,Dai2008}.

IETS spectra taken on a coupled spin system look different from the spectrum recorded on an isolated atom. The STM tip is in fact probing the collective spin excitations of the combined system: the number of possible excitations increases with the number of atoms. Local selection rules as determined by \eref{eq:intensity} determine which excitation appears on which atom. Depending on the coupling strength, different regimes can be discerned. If the coupling is very strong compared to the other energy scales of the problem, spectra recorded on different atoms look exactly the same. This is the case for Mn chains built on Cu$_2$N along the nitrogen row and at the close spacing of 0.36~nm \cite{Hirjibehedin2006}. On the other hand, if the coupling is very small, the spectra look very similar to the single atom, with the coupling acting like a small perturbation \cite{Otte2009,OttePhD}. In the intermediate regime, when the coupling strength is comparable to the anisotropy and the Zeeman term, the spectra look all different to each other but present a collective behaviour, as shown for a ferromagnetically coupled Fe chain \cite{Spinelli2014}. In fact, the first excitation for every atom is at the same energy and has the same intensity, including the outer atoms of the chain that have only one neighbour. For higher energies, the excitations have a different intensity depending on the atom position along the chain, in a manner consistent with spin waves as will be clarified in \sref{section_spinwaves} and \fref{fig:IETS_spin_waves}.

\subsection{Kondo screening}\label{section_Kondo}
The Kondo effect is a many-body interaction between a single magnetic atom and a non-magnetic metal. In the simplest picture (Anderson impurity model \cite{Anderson1961}), the magnetic impurity has only one electron that can be exchanged with the electron bath via a virtual process \cite{hewson1997kondo}. The minimum energy configuration corresponds to the formation of a spin singlet (the Kondo state) between the localized moment and the net spin of the electron bath, that screens the spin of the magnetic impurity. As a result, a sharp resonance peak appears in the electron density of states (DOS) at the Fermi energy.  

Jun Kondo \cite{Kondo1964} used this model to explain a non-conventional behaviour of the electrical resistance, that was observed for the first time during the 1930s \cite{Haas1936}. For a normal metal, the resistance drops with decreasing temperature, and saturates below $\sim10$~K, reaching a value that depends on the concentration of defects in the metal. However, in the 1930s experiment, the resistance of a gold sample was found to decrease with temperature up to a minimum value, while for even lower temperatures it started to increase again. Kondo associated this anomalous increase in the resistivity with the presence of magnetic impurities in the metal and the consequent formation of a Kondo resonant state. This happens below a characteristic temperature, the Kondo temperature $T_K$, that depends on the coupling strength between the impurity and the electron bath \cite{hewson1997kondo}.

The development of experimental tools in nanoscience has opened the doors to new approaches in the study and control of this phenomenon, that before could only be indirectly observed via resistance or magnetic susceptibility measurements. Since 1998 nanotechnologies have made possible to probe a single Kondo impurity: quantum dots \cite{Cronenwett1998,Goldhaber-Gordon1998} and carbon nanotubes \cite{Nygard2000} have allowed to create externally controllable Kondo systems, while the STM has been used to study the interaction between magnetic atoms on top of metal surfaces with atomic resolution \cite{Madhavan1998,Li1998}.

The signature of the Kondo effect in an STM measurement is the appearance of a resonance peak at zero voltage in the IETS spectrum for temperatures lower than $T_K$. When a magnetic atom is deposited directly on top of a metal the interactions with the conduction electrons are very strong, giving rise to a Kondo temperature in the order of 30--100~K \cite{Knorr2002,Wahl2004}. The Kondo interaction, and as a consequence $T_K$, is reduced if there is a decoupling layer like Cu$_2$N separating the magnetic atom from the bulk substrate \cite{Heinrich2004}. For example, for Co atoms on Cu$_2$N, the Kondo temperature is found to be $T_K=2.6\pm0.2$~K \cite{Otte2008}.

In order for a spin to be Kondo screened, a single electron from the bath needs to be able to flip the impurity spin. Therefore, its ground state needs to fulfil certain criteria: it needs to be at least twofold degenerate, and the magnetization $m$ of two degenerate ground states needs to differ by $|\Delta m|=1$. For this reason not all magnetic atoms will be Kondo screened when placed on a metal; known examples observed with an STM are Co \cite{Madhavan1998,Madhavan2001,Knorr2002}, Ce \cite{Li1998}, Ti \cite{Nagaoka2002} and Ni \cite{Jamneala2000} atoms on $\{111\}$ and $\{100\}$ surfaces of Cu, Ag and Au.

Fe on Cu$_2$N has a spin $S=2$ and a negative anisotropy parameter $D$, so the two lowest states are in the subspace of $|m_z\rangle=|2\rangle$ and $|m_z\rangle=|-2\rangle$, which differ by $|\Delta m|=4$. In addition, the presence of the transverse anisotropy term $E$ breaks the degeneracy between the states, causing neither of the two criteria for Kondo screening to be met. Co atoms on the other hand, have a spin $S=3/2$, but $D>0$, so the lowest energy states have $m_z =1/2$ and $m_z =-1/2$. This makes Co a good candidate for Kondo screening. IETS spectra measured on a single Co atom on Cu$_2$N show a sharp resonance peak at zero voltage \cite{Otte2008}.

\section{Magnetic bistability}\label{sec_bistability}
As we have seen, a single magnetic atom may exist in a superposition of two magnetization states despite, or even because of, the presence of strong anisotropy. Yet, bulk magnetic materials consisting of many interacting atoms do become magnetically bistable. The ability of our experiments to control the number of atoms in the system, as well as the anisotropy and the coupling strength, provides a unique opportunity to observe and investigate the emergence of classical magnetic bistability with increasing atom number.

In systems with easy axis anisotropy, two metastable energy minima are separated by an energy barrier providing stability to two opposite spin orientations \cite{Skomski2003,Chudnovskiy2014}. Several effects may work to reduce this stability. First, thermal excitations may overcome the barrier, or a strong enough external magnetic field may make the barrier sufficiently asymmetric to destabilize one side. Also electron-induced excitations (caused e.g. by interactions with substrate electrons or tunnelling electrons in the case of an STM experiment) can allow the system to gain energy and climb the ladder of states to reach the other side. But even at very low temperatures and in the absence of a magnetic field or electron excitations, quantum tunnelling of magnetization can allow the system to go directly from one state to the other in case finite overlap between the states exist \cite{Gatteschi2003,Leuenberger2001}.

With the miniaturization of memory storage devices reaching levels where quantum effects can no longer be neglected, control of the magnetic stability at the atomic scale is becoming a topic of increasing interest. Bistable behaviour has been reported in several systems composed of only several atoms, e.g. Fe nanoislands on W(100) \cite{Krause2009,Herzog2010} and Co chains on Pt \cite{Gambardella2002}. More recently, spin-polarized STM experiments have revealed switching between magnetic states on spin structures built through atom manipulation \cite{Loth2012,KhajetooriansScience2013}, and have shed light on the spin dynamics during a magnetic switch \cite{Spinelli2014}. In 2013 bistability was reported on individual rare-earth atoms deposited directly on a metal \cite{Miyamachi2013}. Very recently it was demonstrated by means of electronic pump probe spectroscopy \cite{LothScience2010} that stability in small structures built on Cu$_2$N can be modified via exchange interaction with the tip \cite{Yan2014}.

In the example of Loth et al. (2012) \cite{Loth2012}, construction of antiferromagnetically coupled Fe chains on Cu$_2$N/Cu(100) resulted in bistability. By creating eight arrays of $(2\times6)$ Fe atoms, the authors were able to build the first atomic byte and store information in it for hours, providing a significant advancement towards the creation of nanoscale memory storage devices. In the case of an antiferromagnet, the two lowest energy states are the two classical N\'{e}el states, in which the spins are counter-aligned to form a collective state with zero net magnetization. In this experiment it was observed current and temperature induced switching between the N\'{e}el states. Moreover, the intrinsic (i.e. temperature independent) switching rate, that the authors attribute to quantum tunnelling of magnetization, was found to depend very sensitively on the number of atoms in the structure: by adding just two atoms to a six-atom chain, the intrinsic switching rate reduced by three orders of magnitude. 

An antiferromagnetic system is a relatively complex magnetic object: technically, neither of the N\'{e}el states is ever an eigenstate of the system. In order to better understand the spin dynamics during a magnetization reversal, it may therefore be easier to study a ferromagnetic system, as shown in Ref.~\cite{KhajetooriansScience2013}. Here, the authors studied the magnetic switching of a cluster of Fe atoms, built directly on a metal Cu(111) substrate, ferromagnetically coupled to each other. Unfortunately, due to the short-range coupling between the iron atoms, this system could not be probed or excited with single spin resolution. In Ref.~\cite{Spinelli2014}, bistability in ferromagnetically coupled Fe structures on Cu$_2$N, enabling larger atom separation, was achieved.

In \sref{section_detection} we will present methods to detect magnetic switching, including arrangements to measure regimes that are too fast or too slow for conventional detection. Next, in \sref{section_spinwaves} we will investigate the role played by stationary spin waves, confined into a ferromagnetic Fe chain, to revert its magnetization. Finally, in \sref{section_backaction} we will present novel switching data on antiferromagnetic structures that may be used to extract information about the spin polarized the STM tip.

\subsection{Telegraph noise detection}\label{section_detection}
\begin{figure}[htb!]
\centering
\includegraphics{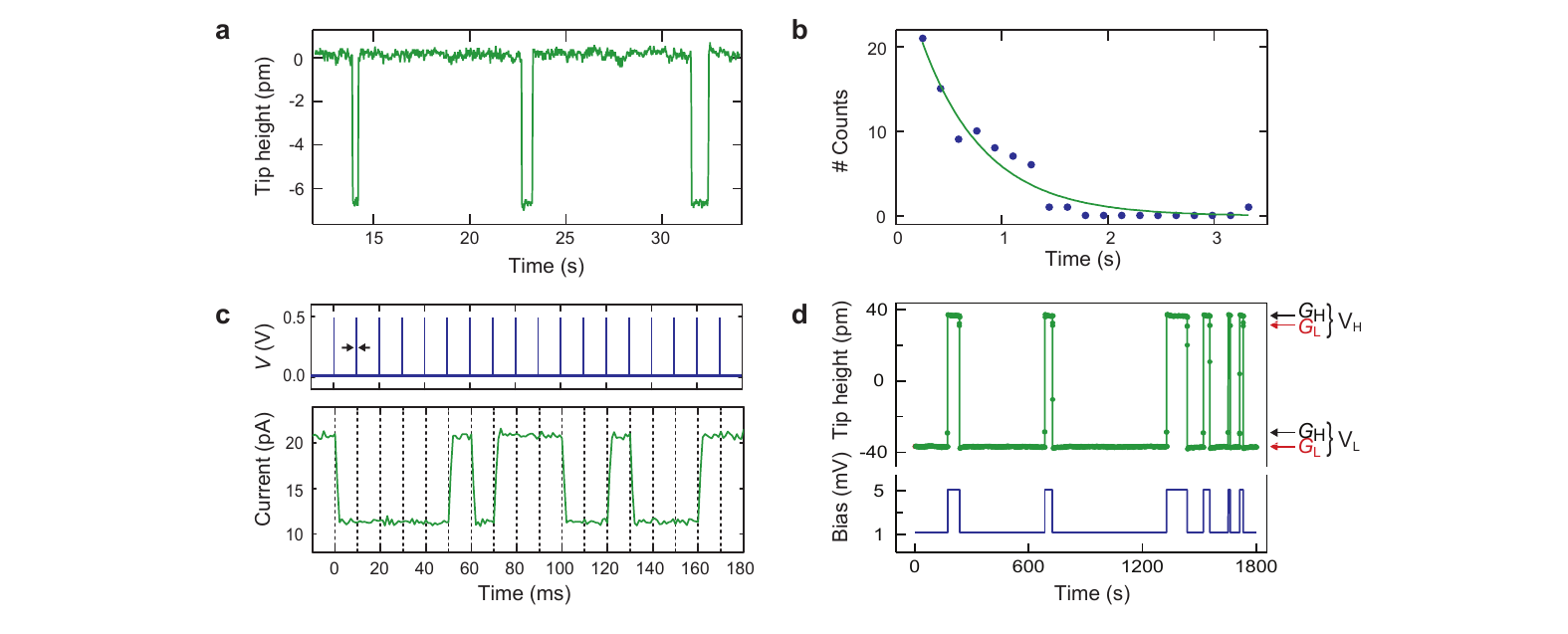}
\caption{(a) Part of a constant-current switching trace, recorded for the second atom of a chain composed of 6 Fe atoms ferromagnetically coupled to each other. It was recorded with sample bias voltage $V=3.7$~mV (above the 3.5~mV threshold for inelastic excitations), current $I=20$~pA, magnetic field $B=200$~mT and temperature $T=330$~mK. (b) Time histogram (blue markers) corresponding to the full switching trace from $G_{\rm L}$ to $G_{\rm H}$ of panel (a). From the exponential fit (green curve), the lifetime of the low current state can be extracted as the decay time. (c) Example of a pulsed measurement, taken from \cite{Loth2012}. The top part shows the voltage pulses, $10$~ns long, and below the corresponding response of the system. (d) Feedback mechanism to measure very slow switching rates, developed in \cite{Spinelli2014}. As soon as a switch from $G_{\rm H}$ to $G_{\rm L}$ is detected in the $V_{\rm H}$ state, the bias is suddenly lowered, with the system still in $G_{\rm L}$ but at low bias (below threshold). When the system switches back to $G_{\rm H}$, the bias voltage is set back to the high value.}
\label{Fig:telegraph_noise}
\end{figure}

Magnetic bistability can be detected by making the magnetic structure interact with a spin polarized STM tip. In the resulting magnetic tunnel junction, the measured conductance will be higher when the spins of tip and sample are in parallel alignment to each other and lower when they are antiparallel. If an atom or a structure is switching between two magnetization states, the tip will measure a current (and therefore conductance) that is alternating stochastically between two different values, corresponding to the two different configurations, that we call high-current $G_{\rm H}$ and low-current $G_{\rm L}$. If the measurement is performed in constant-current mode, the switching behaviour of the current can be observed as a switching behaviour of the tip height. An example trace of this telegraphic noise signal is shown in \fref{Fig:telegraph_noise}a.  

In order to estimate the switching rates between the two states, those telegraph noise curves need to be recorded for enough time to detect a statistical number of switches. Between switches, the system spends a time interval in a specific state; by making a histogram of the different time intervals spent by the system in one state, we observe an exponential decay. We can at this point extrapolate the Poissonian lifetime of each state from the two exponential decay times, and the inverse of the lifetimes are a measurement of the switching rates. This procedure is shown in \fref{Fig:telegraph_noise}b.

When the switching rate exceeds the bandwidth of the current amplifier, for example when a small structure is measured or when a larger bias voltage is applied, a pulsed measurement scheme can be used as shown in \fref{Fig:telegraph_noise}c \cite{Loth2012,Spinelli2014}. The idea is to apply a voltage larger than the threshold for inelastic excitations in very short pulses, on top of a DC voltage below the threshold. It is assumed that the probability for a switch to occur in the low-voltage intervals is zero. In these time intervals the magnetization can be considered frozen. As a result, one can say that the amount of relevant time passed in a time interval $\Delta t$ is $\Delta t (t_1/(t_0+t_1))$, where $t_1$ and $t_0$ are the pulse length and the zero voltage interval respectively. In other words, the measurement time has slowed down with respect to real time by a factor $t_1/(t_0+t_1)$, allowing much faster switching processes to be investigated.

On the other hand, in some cases the switching rate can be extremely slow. If the system is a ferromagnet, it will spend most of its time in one state, even if the applied field (required for magnetizing the SP tip) is very small. As a consequence, in order to obtain enough statistics to extract lifetimes, a way to speed up the process needs to be found. An example is given in Ref.~\cite{Spinelli2014}, in which a feedback mechanism is used as shown in \fref{Fig:telegraph_noise}d. If the system spends most of its time in state $G_{\rm H}$, the switching rate from $G_{\rm H}$ to $G_{\rm L}$ is very slow. So to speed up the measurement, the (slow) rate from $G_{\rm H}$ to $G_{\rm L}$ is measured at a high voltage $V_{\rm H}$ (above inelastic excitation threshold), while the faster rate from $G_{\rm L}$ to $G_{\rm H}$ is measured at low voltage $V_{\rm L}$ (below threshold). As soon as a switch from $G_{\rm H}$ to $G_{\rm L}$ in the high voltage state $V_{\rm H}$ is recorded, the voltage is suddenly reduced to $V_{\rm L}$, with the system ending in state $G_{\rm L}$ at low bias voltage. Then, when a switch to $G_{\rm H}$ is detected, the voltage is restored to $V_{\rm H}$. This method enables measurement of the intrinsic switching rate from the unfavoured to the favoured state, without having to wait endlessly while the system spends time in the favoured state.

\subsection{Spin wave mediated reversal}\label{section_spinwaves}
Very recently, in Ref.~\cite{Spinelli2014} real-space atomic scale imaging of standing spin waves was reported. The spin waves were confined inside an artificial ferromagnet containing only six Fe atoms. In this experiment, the possibility to have ferromagnetic coupling between Fe atoms over relatively large distance on Cu$_2$N \cite{Bryant2013}, enabled to study the collective excitations of the structure while addressing each atom individually. By using a combination of spin polarized STM to detect the magnetic switching, and IETS to characterize the inelastic excitations, the authors were able to study the complex spin dynamics that governs the magnetization reversal of a ferromagnet. In this section we will focus on the role played by spin waves during this process.  

\begin{figure}[htb!]
\centering
\includegraphics{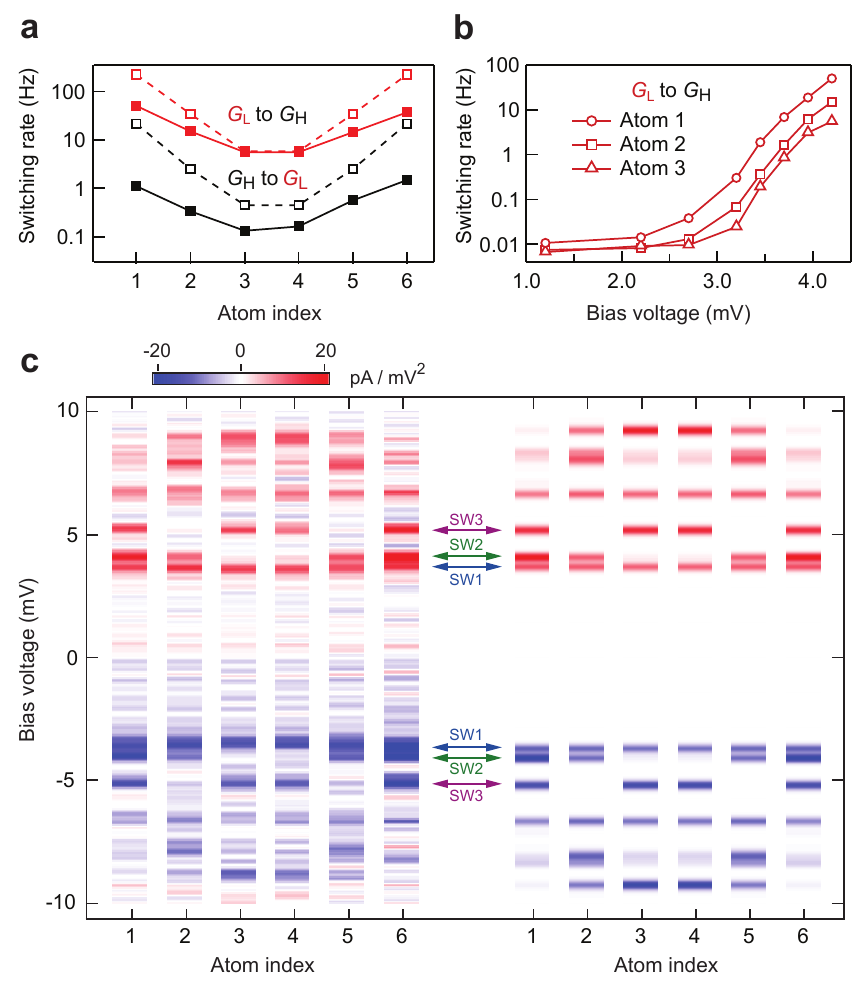}
\caption{(a) Telegraph noise switching rates in both directions plotted as a function of position along the chain. Filled markers connected with solid line are the experimental data, open markers connected with dashed line are the theoretical simulations. For this measurement: sample bias voltage $V=4.2$~mV, magnetic field $B=200$~mT and temperature $T=330$~mK. (b) Measured switching rates from the low conductance state $G_{\rm L}$ to the high conductance $G_{\rm H}$ as a function of the applied voltage, for the first three atoms of the chain. Magnetic field and temperature same as (a). (c) IETS spectra (d$^2I$/d$V^2$) as a function of position along the chain, at $B=200$~mT and $T=330$~mK. The left panel shows the experimental data; the right side the corresponding simulations. The three lowest energy excitations, that are confined spin waves, are identified. All panels reproduced from Ref.~\cite{Spinelli2014}.}
\label{fig:IETS_spin_waves}
\end{figure}

On a ferromagnetic Fe chain, the switching rates measured along the chain are not constant. The observed spatial dependence of the switching rates of a six-atom chain is shown in \fref{fig:IETS_spin_waves}a for a specific value of applied voltage. Not only is the switching fastest on the outer atoms (atoms 1 and 6), also atoms 2 and 5 have a rate that is significantly higher than the rate found on atoms 3 and 4. As such, explanations based on the fact that the outer atoms, unlike the inner atoms, have only one neighbour and are therefore easier to switch can be discarded. Also, a strong dependence on the applied voltage is observed (\fref{fig:IETS_spin_waves}b): when the applied voltage is $V\sim3.5$~mV, the switching rates rapidly increase for all the atoms in the chain.  

To shed some light into this behaviour, IETS spectra were measured with a non polarized tip, as a function of atom position along the chain. Recorded spectra and corresponding simulations are shown in \fref{fig:IETS_spin_waves}c. Instead of the usual d$I$/d$V$, here the color scale shows its derivative (d$^2I$/d$V^2$): the spin excitations appear as peaks (dips) for positive (negative) sample voltage, at the same energies of the steps in the corresponding differential conductance curves. Surprisingly, the first excitation occurs at the same energy along the chain and has the same intensity for all atoms. Apparently these are truly collective excitations of the chain and no difference is observed between atoms having one neighbour (atoms 1 and 6) and those having two. The second excited state, at an energy very close to the first, and the third have both a nodal modulation along the chain: they are less intense on the center atoms and on atoms 2 and 5 respectively. These three low energy states are identified as standing spin waves confined in the chain. 

The dependence of the switching rate on the bias voltage can be explained by noting that the value at which the rates start to rise corresponds to the inelastic excitation threshold: the tip is driving the switching by excitations above the energy barrier. However, this first spin wave state SW1 is homogeneous along the chain, so it cannot explain the observed spatial dependence. But the second excited state SW2 is very close in energy, and with theoretical models based on Pauli master equations, it can be shown that there is a non negligible probability that, once the system is excited in SW1 by the tip, a chain of subsequent excitations occurs, temporarily occupying SW2 (that is responsible for the spatial modulation) and ending up on the other side of the barrier. So, to controllably revert the magnetization of a ferromagnet, the easiest way is by exciting it into a spin wave state with a node on the center, that can assist the switching \cite{Spinelli2014,Rohart2014}.       
  
\subsection{Back-action onto the tip}\label{section_backaction}
In most experiments involving spin-polarized STM, the tip is considered to be either permanently magnetized or, when polarization is achieved by picking up one or a few magnetic atoms, paramagnetic and thereby constantly oriented in the direction of an external magnetic field. Considering the low coordination of atoms at the apex of a tip, however, the assumption of static polarization is likely not very realistic. In this section we will show novel data on magnetic switching that can be used to extract information about the STM tip itself.

Not much it is known about the few atoms close to the tip apex that are actually taking part in the tunnelling process. Whereas the bulk material of the tip is typically a hard metal (e.g. PtIr), in order to make it atomically sharp, it is regularly indented in the softer metal substrate. In our experiments the sample is a copper crystal, so we can imagine our tip to be copper coated. In order to polarize the tip, a single magnetic (e.g. Fe) atom is picked up to replace the apex Cu atom. This stage cannot be controlled very well: sometimes during manipulation Fe atoms are picked up but cannot be released afterwards. Therefore, the number of Fe atoms on the tip and how they are exactly positioned is unknown. Consequently, the manner in which the tip will respond to its environment, e.g. an external magnetic field or a stray field induced by the sample, is not predictable. Due to interactions acting back onto the tip, the observed switching behaviour of a bistable structure detected by the tip may be non-uniform along the structure. As demonstrated below, these variations can be used to extract information about the tip apex.  

\begin{figure}[htb!]
 \centering
 \includegraphics{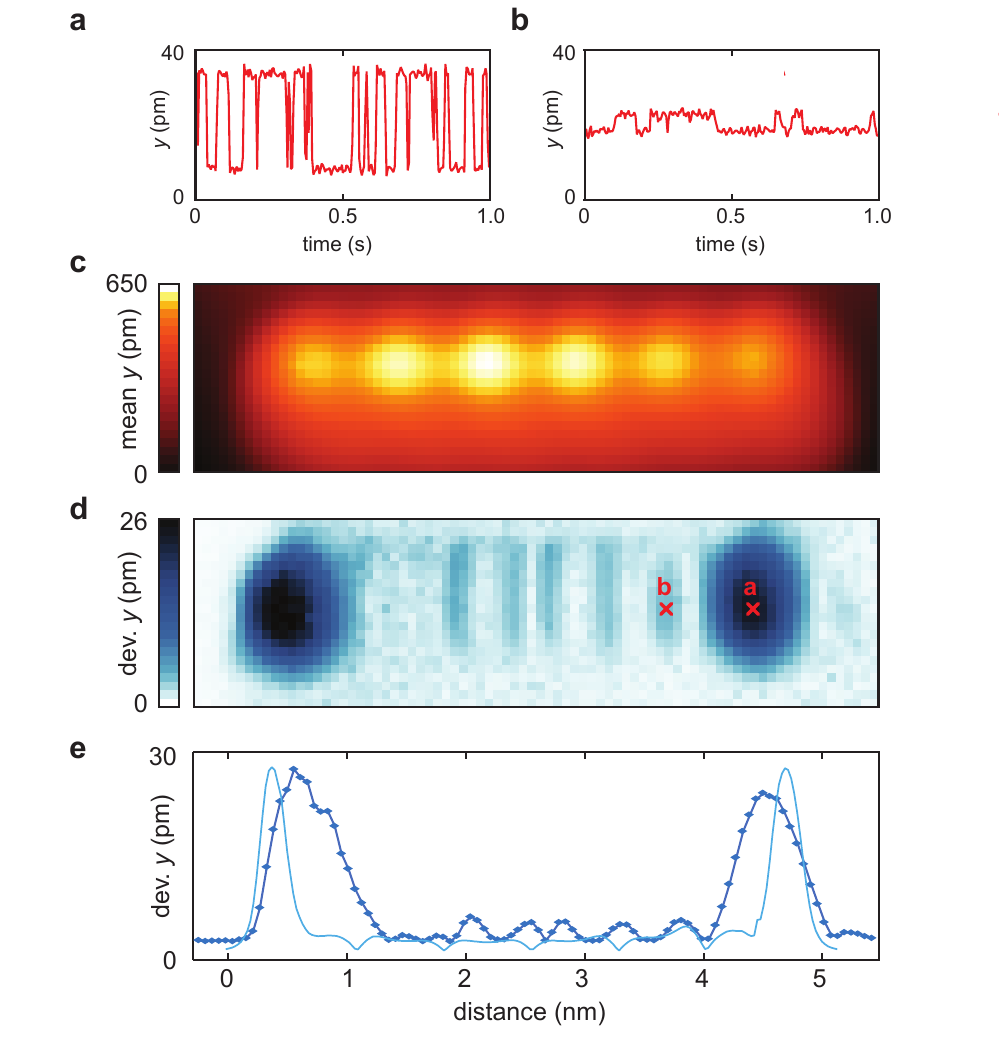}
 \caption{\label{fig:tip}(a)-(b) Two examples of telegraph noise traces measured, respectively, for an outer and an inner atom of a chain composed of 6 Fe atoms antiferromagnetically coupled to each other. Exact locations where those traces were recorded is shown in (d). (c) Mean value of the measured tip height (i.e. topography) as a function of the spatial position. (d) Standard deviation of the height distribution along the chain, showing switching amplitudes as a function of position. (e) Line-cut of (d) along the center of the chain (marks connected with dark blue line), showing the variations in tip height along when the tip is moving from atom to atom. It is interesting to note the ratio of the peaks for outer and inner atoms and the structure in correspondence of the inner atoms. The (thinner) light blue line is the simulated curve, based on the model described in \sref{section_backaction}.}
\end{figure}

At the beginning of this section, we have introduced the antiferromagnetic Fe structures exhibiting bistable behaviour, measured in Ref.~\cite{Loth2012}. We have reproduced this experiment obtaining similar results concerning switching rates. However, in some cases we observed a spatial variation in the switching amplitude not reported before. In particular, we found two remarkable facts. First, the difference in conductance between the two N\'{e}el states (and consequently the tip height difference $\Delta y$) was more pronounced for the outer atoms of the chain than for all the others; two examples of those different amplitude telegraph noise switching traces are shown in figures~\ref{fig:tip}a--b. Second, for the inner atoms of the chain, a subatomic periodicity was found: the largest switching amplitude was not measured on top of the atoms but slightly off either side of the atom. A measurement of $\Delta y$ as a function of position along the chain is presented in figures~\ref{fig:tip}d--e, for a chain with six Fe atoms. As can be seen by the density of experimental points, several data points at fixed distance within each other have been acquired per atom. This allows to see how the switching traces vary while the tip is moving across each atom of the chain. For reference, in \fref{fig:tip}c the topography, measured simultaneously with the switching data, is presented.

As discussed in \sref{section_SP-STM}, in a spin-polarized STM measurement, the tunnel current can be expressed as the non polarized component times a term that depends on the product between tip and sample polarizations and the cosine of the angle $\beta$ between the two magnetization directions \eref{eq:I_SP}. For the chain studied here, in the ground state the easy-axis anisotropy of the Fe atoms on Cu$_2$N aligns the spins along the direction in which the chain is built (defined as $z$-direction) \cite{Loth2012}. At zero magnetic field this results in two degenerate ground states (the N\'{e}el states) with alternating spin orientations. By restricting our attention only to the two N\'{e}el states and magnetic fields applied along $z$, the angles between the spin polarization of the tip and the $j$-atom in the chain can only be $\beta_j=0$ or $\beta_j=\pi$. So the spin polarized current of one N\'{e}el state becomes ($P_{j}^{S}$ is the polarization of the $j$-th atom of the chain):

\begin{equation}\label{eq:I_SP_Neel}
I_{\rm SP}(\mathbf{d})\propto \sum_{j=1}^{n} I_0(\mathbf{d}) \left[1+\left(-1\right)^{j}P^{T}P_{j}^{S}\right].
\end{equation}

The expression for the current of the other N\'{e}el state will be the same but with inverted atom polarizations (i.e. substitute $(-1)^{j}$ with $(-1)^{j+1}$ in \eref{eq:I_SP_Neel}). 

Here we present a model based on the notion that the magnetic atom located at the tip apex can pivot around the apex. The magnetic atom is represented as a pendulum subject to an harmonic potential, that is influenced both by the external magnetic magnetic field and the stray field induced by the magnetic structure on the surface. As a result, the chain switching behaviour detected by the tip will be non uniform along the structure. A similar model has been proposed by Hapala and coworkers \cite{HapalaPRL2014} to explain the high-resolution molecular imaging of molecules by means of IETS described in Ref.~\cite{Chiang2014}.

We assume that the atomic spins of the antiferromagnetic chain will generate a stray field, that can be modelled as the total magnetic field created by the magnetic dipole moments $\boldsymbol{\mu}_j$, positioned at each atom site $\mathbf{r}_j$. Each dipole will generate a magnetic field at position $\mathbf{r}$, $\mathbf{B}_{j}(\mathbf{r})$, that can be written as \cite{Griffiths_ED}:

\begin{equation}
\mathbf{B}_{j}(\mathbf{r})=\frac{\mu_{0}}{4\pi\left|\mathbf{d}\right|^{3}} \left[3\left(\boldsymbol{\mu}_{j}\cdot\hat{\mathbf{d}}\right)\hat{\mathbf{d}}-\boldsymbol{\mu}_j\right] +\frac{2\mu_{0}}{3}\boldsymbol{\mu}_{j}\delta^{3}(\mathbf{d}).
\end{equation}  

In the previous equation, $\mu_0$ is the vacuum permeability, $\delta^3(\mathbf{d})$ is the three-dimensional Dirac delta function, $\mathbf{d}\equiv(\mathbf{r}-\mathbf{r}_j)$ is the relative distance between the position where the field is measured $\mathbf{r}$ and the position $\mathbf{r}_j$ of the magnetic moment $\boldsymbol{\mu}_{j}$ (where $\hat{\mathbf{d}}$ denotes the unit vector in the direction of ${\mathbf d}$). Since all the dipole moments of the chain are assumed to be oriented along the $z$-axis and the Fe atoms on Cu$_2$N have spin $S=2$, we can write $\boldsymbol{\mu}_{j}=\pm(-1)^{j+1}2g\mu_{\rm B}\hat{z}$, with the $\pm$ sign identifying one of the two possible N\'{e}el states. The total stray field generated by the structure will then be the sum of all the dipole moments contributions:

\begin{equation}
\mathbf{B}(\mathbf{r})=\sum_{j=1}^{n}\mathbf{B}_{j}(\mathbf{r})\label{eq:stray_field}.
\end{equation}  

The calculated stray field \eref{eq:stray_field} felt by the tip at a typical tip-sample distance of $3$~{\AA} is a few mT, much smaller than typical values of external magnetic field applied during the experiment. However, a non-uniform field, like the one generated by the Fe chain, will also exert a force on any magnetic dipole, like the dipole moment $\boldsymbol{\mu}_{\rm apex}$ associated to the spin of the apex atom of the spin polarized tip. The force is given by:

\begin{equation}
\mathbf{F}_{\rm Dipole}=\nabla\left(\boldsymbol{\mu}_{\rm apex}\cdot\mathbf{B}\right),
\end{equation}

where $\nabla\equiv[({\partial}/{\partial x})\hat{x} +({\partial}/{\partial y})\hat{y}+ ({\partial}/{\partial z})\hat{z}]$ is the gradient operator and $\boldsymbol{\mu}_{\rm apex}=-2g\mu_{\rm B}\hat{z}$ if the external field is oriented along the negative $z$-direction. 

When the chain switches from one N\'{e}el state to the other, the stray magnetic field \eref{eq:stray_field} changes sign, and this leads to $\mathbf{F}_{\rm Dipole} \rightarrow -\mathbf{F}_{\rm Dipole}$. The external magnetic field, on the other hand, is uniform and thereby does not exert any force on the tip apex atom. As such, the potential energy of the tip due to the magnetic field is given by (with $B_z$ the $z$-component of the stray field):

\begin{equation}
U_{\rm Dipole}=-\mu_{\rm apex}B_z.
\end{equation}

\begin{figure}[htb!]
 \centering
 \includegraphics{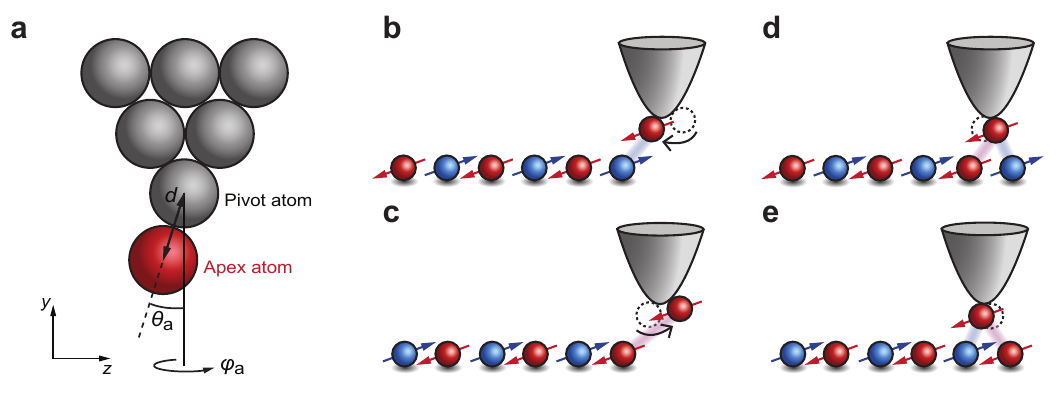}
 \caption{\label{fig:tip_model} (a) Schematic of the last few atoms of the tip according to the pendulum-tip model. (b)-(e) Illustration of the motion of the tip apex in response of the stray magnetic field generated by the structure in both N\'{e}el states. Near the edge of the chain (b,c), the motion will be much more dramatic than over the center of the chain (d,e), where the apex atom feels conflicting influences from neighbouring atoms in the structure.}
\end{figure}

The displacement of the magnetic apex atom due to the stray field generated by the structure may affect the switching signal as follows. Let's assume the apex atom is bound to a single non magnetic atom of the tip (see \fref{fig:tip_model}a), that we call the pivot atom. We treat those two atoms as hard spheres, with a fixed distance $d=1$~{\AA} between their respective centres. The apex atom is allowed to rotate in the two directions identified with the angles $\phi_a$ and $\theta_a$. If we treat this system as a simple harmonic oscillator, the apex atom will have, at position $\theta_a$, a potential energy given by $\frac{1}{2}K\theta_a^2$. Furthermore, we restrict the motion of the atom to $0\leq\theta_a\leq\pi/4$.  

The total potential energy felt by the apex atom will be a function of the position of the pivot-atom above the chain $\mathbf{r}_{\rm T}$ and the spherical coordinates of the apex-atom with respect to the pivot $\left(\phi_a,\theta_a\right)$:

\begin{equation}\label{eq:pendulum}
V(\mathbf{r}_{\rm T},\theta_a,\phi_a)=\frac{1}{2}K\theta_a^{2}-\mu_{\rm apex}B_z(y_{\rm T},z_{\rm T},\theta_a,\phi_a).
\end{equation}

The equilibrium position of the apex atom can be determined by numerical minimization of \eref{eq:pendulum} with respect to the spherical coordinates $\left(\phi_a,\theta_a\right)$ for a fixed pivot position $\mathbf{r}_{\rm T}$  above the chain. This leads to an equilibrium position $\left(\phi_{\rm eq},\theta_{\rm eq}\right)$ from which the apex position $\mathbf{r}_{\rm apex}$ above the chain can be determined. Once this position is known, we can calculate the corresponding polarized tunnel current for the two N\'{e}el states by using \eref{eq:I_SP_Neel}, and therefore their difference. Performing this algorithm for a range of tip positions along the chain $z_{\rm T}$ and different tip height $y_{\rm T}$, we can simulate the switching data as shown in \fref{fig:tip}e. 

In this model the work function $\varphi$ of the tunnel barrier was used as a fitting parameter. This quantity enters in the non-polarized current \eref{eq:I} via the following relation:

\begin{equation}
\varphi=\frac{\hbar^2\kappa^2}{2m},
\end{equation}

with $\kappa$ being the decay rate of the tunnel current with increasing distance, $m$ the electron mass and $\hbar$ the reduced Planck constant. From the fits, we have obtained a work function of $\varphi=3.8$~eV, very close to the typical value of a tunnel junction $\varphi=4.2$~eV.

The simulated switching curve show qualitative agreement with the measured data, e.g. the large relative peak height of the outer atoms compared to the inner atoms of the chain and the appearance of a subatomic double periodicity on the inner atoms. Figures~\ref{fig:tip_model}b--e show an interpretation of the motion of the apex atom when it is subject to the force due to the chain stray field $\mathbf{F}_{\rm Dipole}$ in the proposed model. The difference in peak height between inner and outer atoms comes out naturally when assuming a spin polarized current tunnelling not only into the atom straight underneath the tip but also diagonally into the two neighbouring atoms, with opposite magnetization direction. The effect of the signal coming from those two atoms cancels out part of the spin polarized current, and this reduces the apparent tip height variation. This effect will be smaller when tunnelling into the outer atoms that have only one neighbour. No pivoting atom is needed to explain this observation. The double periodicity found on the inner atoms of the chain, however, cannot be reproduced without introducing the stray field induced by the structure on the pivot atom.

In this way we have shown how, starting from switching data measured on a bistable chain, we can gain information on the functionalized spin-polarized tip, that is otherwise considered an unknown parameter in most of STM experiments. We note that the effect described in this chapter may be related to what was very recently shown in Ref.~\cite{Yan2014}, in which the magnetic structure interacts with the STM tip via exchange interaction.

\section{Conclusions and outlook}
In this paper we have reviewed the state of the art of the atomic scale magnetism studied by means of a scanning tunneling microscope. Low dimensional magnetism is a topic that has attracted a lot of theoretical interest, starting form the early works of Ising \cite{Ising}, Bethe \cite{Bethe} and Heisenberg \cite{Heisenberg} at the end of the 1920s. Only very recently, it has been possible to experimentally build magnetic lattices, using vertical atom manipulation, and probe them locally. This is a very promising avenue for realizing quantum mechanical systems and study their transitions to the classical limit.

After reviewing the principal experimental techniques needed for this fascinating study, in this work we have summarized the principal aspects of the physics of single spins. When an atom is embedded in a surface, it can develop a magnetocrystalline anisotropy, that is a result of the interplay between the spin-orbit coupling and the crystal field splitting induced by the broken symmetries of the surface. A magnetic impurity on a surface can also be Kondo-screened by the substrate electrons and it can react to an externally applied magnetic field, due to the Zeeman effect. Finally, two magnetic atoms can interact with each other, and the strength and sign of their interaction will depend on their relative distance. Here we have shown how all those physical properties can not only be probed but also tuned by adjusting the exact position of an atom (or of few atoms) on a substrate.

All those effects can give rise to magnetic bistability. Very recently a few examples of systems composed only of a few atoms have been proven to be bistable, opening the way to new approaches towards the realization of nanoscale magnetic storage devices. Next to showing some techniques to actually detect this bistability, even in cases for which conventional measurements are not sufficient, we have given an insight on the role played by standing spin waves in order to revert the magnetization of a ferromagnet composed of only few atoms. Finally, we have shown how data collected on a bistable structure could be used to extract useful information on the apex of an STM tip, the part that is actually involved in the tunnelling process.

Now that structures with ferromagnetic and antiferromagnetic interactions can be atomically engineered, the way is paved for experiments in which we can look at the evolution of spinons and magnons in one- and two-dimensional lattices. Unlike other techniques used to study quantum magnetism, e.g. utracold atoms, the field of atomic engineering enables to make customized modification to the lattice, like local substitutional doping, defects injection or magnetic frustration. We foresee that this approach is reaching a state of maturity where we expect a synergy to be created between experimental atomic surface science and theoretical quantum magnetism.

\section*{References}
\bibliography{library}

\end{document}